\documentstyle[12pt,epsfig]{article}
\newlength{\dinwidth}
\newlength{\dinmargin}
\newcommand{\ba}{\begin{array}}
\newcommand{\ea}{\end{array}}
\newcommand{\beq}{\begin{equation}}
\newcommand{\eeq}{\end{equation}}
\newcommand{\bea}{\begin{eqnarray}}
\newcommand{\eea}{\end{eqnarray}}

\def\parallel{| \hskip-0.03cm |}

\def\bce{\begin{center}}
\def\ece{\end{center}}

\begin{document}
\thispagestyle{empty}
\addtocounter{page}{-1}
\begin{flushright}
SNUST 99-005\\
KIAS-P99034\\
{\tt hep-th/9905205}\\
\end{flushright}
\vspace*{1.3cm}
\centerline{\Large \bf Thermodynamics of Large-N Super Yang-Mills Theory}
\vskip0.3cm
\centerline{\Large \bf and}
\vskip0.3cm
\centerline{\Large \bf  AdS/CFT Correspondence~\footnote{Work supported in 
part by KOSEF Interdisciplinary Research Grant 98-0702-07-01-5 and 
SRC Program, KRF International Collaboration Grant, Ministry of Education 
Grant 98-015-D00054, and The Korea Foundation for Advanced Studies Faculty 
Fellowship. }}
\vspace*{1.5cm} 
\centerline{\large \bf Chanju Kim${}^a$ {\rm and} Soo-Jong Rey${}^{b,c}$}
\vspace*{1.0cm}
\centerline{\large\it ${}^a$ School of Physics, 
Korea Institute for Advanced Study, Seoul 702-701 Korea}
\vskip0.3cm
\centerline{\large\it ${}^b$ Physics Department, Seoul National University,
Seoul 151-742 Korea}
\vskip0.3cm
\centerline{\large\it ${}^c$ Asia-Pacific Center for Theoretical Physics,
Seoul 130-012 Korea}
\vspace*{0.8cm}
\centerline{\tt cjkim@ns.kias.re.kr, \hskip0.5cm sjrey@gravity.snu.ac.kr}
\vskip2cm
\centerline{\Large\bf abstract}
\vspace*{0.5cm}
Thermodynamics of $d=4$, ${\cal N}=4$ supersymmetric $SU(N)$ Yang-Mills theory 
is studied with particular attention on perturbative expansion at weak
`t Hooft coupling regime and interpolation to strong coupling regime thereof.
Non-ideal gas effect to free-energy is calculated and found that leading- 
and next-to-leading-order corrections contribute with relative opposite signs.
Pad\'e approximant method is adopted to improve fixed-order, perturbative 
series and is found to decrease free-energy monotonically as `t Hooft coupling 
parameter is increased. This may be regarded as an indication of smooth 
interpolation of thermodynamics between weak and strong `t Hooft coupling 
regimes, as suggested by Maldacena's AdS/CFT correspondence. 

\vspace*{1.1cm}

\baselineskip=18pt
\newpage

\section{Introduction}
Maldacena's proposal \cite{maldacena} relates large $N$ limit of the $d=4$, 
${\cal N}=4$ supersymmetric $SU(N)$ Yang-Mills theory to Type IIB 
superstring theory on 
$AdS_5 \times S^5$. According to the correspondence, parameters of Yang-Mills 
and string theories are related each other as:
\bea
g^2_{\rm YM} = 2 \pi g_{\rm st} \, , 
\qquad g^2_{\rm YM} N   = {1 \over 2} {R^4 \over \alpha'^2},
\nonumber
\eea
where 
$R$ is the curvature radii of both $AdS_5$ and $S^5$, and $g_{\rm st}$ and 
$\alpha'$ are string coupling parameter and string slope parameter, 
respectively. Exploration of the proposed duality over the entire parameter 
space $(g^2_{\rm YM}, N)$ is seemingly a formidable, if not impossible, task.
However, the above relation already implies that physics of super Yang-Mills
theory in the strong `t Hooft coupling limit, $\lambda^2 \equiv 
g^2_{\rm YM} N \rightarrow \infty$, is directly accessible from systematic 
expansion in the zero-slope limit, $(\alpha' / R^2) \rightarrow 0$, of Type 
IIB string theory, viz. gauged $AdS_5$ supergravity. 
Indeed, all the checkpoints regarding the duality 
examined so far have been exclusively in this limit.
It is therefore of interest to explore, for various physical quantities, 
the Maldacena's proposal beyond the leading order in strong `t Hooft coupling
in super Yang-Mills theory and in $(\alpha' / R^2)$ in gauged 
$AdS_5$ supergravity. Of particular interest is whether the physical quantity
is an analytic function over the entire parameter space. If so, it would be 
interpolating smoothly between weak and strong coupling regimes. 
In this paper, as an example of such physical quantities, we would like to 
investigate thermodynamic free-energy density $F(N, \lambda)$, with 
particular attention to the issue of interpolation between weak and strong 
`t Hooft coupling regimes. The issue is particularly relevant, as, based on
certain assumptions on analyticity and convergence of the perturbative 
series, it has been claimed that such a smooth interpolation is not possible
\cite{li}. 

In the strong `t Hooft coupling limit, the free-energy density has been 
calculated in \cite{gubserklebanovtseytlin}, including leading-order 
correction. Their calculation goes as follows. According to the Maldacena's 
proposal extended to finite temperature, super Yang-Mills 
theory at high temperature $T$ is described, in large $N$ and strong `t 
Hooft coupling limit, by a Schwarzschild $AdS_5$ black hole \cite{hawkingpage} 
at Hawking temperature $T$. The metric of the black hole is given by
\bea
ds^2 = 
{{\rm U}^2 \over R^2} \left(- f({\rm U}) dt^2 + d {\bf x}^2_{\parallel} \right)
+
{R^2 \over {\rm U}^2} \left(f^{-1} ({\rm U}) dr^2 
+ {\rm U}^2 d \Omega_5^2 \right),
\label{metric}
\eea
where, expressed in terms of Yang-Mills paramters, the Schwarzschild 
factor $f({\rm U})$ is given by
\bea
f({\rm U}) = \left( 1 - {{\rm U}_0^4 \over {\rm U}^4} \right) \, , 
\qquad {\rm U}_0 = 2 \pi \lambda T \, .
\nonumber
\eea
As shown in \cite{gubserklebanovtseytlin}, the metric Eq.(\ref{metric}) turns
out to be a solution of gauged $AdS_5$ supergravity equations of motion not 
only at leading-order (Einstein-Hilbert action) but also including ${\cal O} 
(\alpha'^3)$ string correction terms. The free-energy density $F$ is then
obtained from the Euclidean supergravity action of the black hole times 
Hawking temperature $T$. The result is
\bea
F(N, T) &=& h(\lambda) \left( - {1 \over 6} \pi^2 N^2 T^4 \right)
\nonumber \\
h(\lambda) &=& \left[ {3 \over 4} + {45 \over 32} \zeta(3) \lambda^{-3/2}
+ \cdots \right] \, ,
\label{free}
\eea
and hence indicates that the free-energy density is diminished by a factor
$h(\lambda) \sim 3/4$ compared to the free-energy of ideal gas of 
${\cal N}=4$ super Yang-Mills theory, $- \pi^2/6 \cdot N^2 T^4$. 
Branch-cut of the free-energy density at $\lambda = \infty$ is a genuine 
prediction for the strong coupling limit, along with the heavy quark potential,
both at zero \cite{reyyee, maldacenawilson}
and finite temperatures \cite{witten, reytheisenyee, brandhuber}, and 
anomalous dimensions of long supermultiplets \cite{gubserklebanovpolyakov, 
wittenhol}. 

Were it interpolating smoothly over the entire `t Hooft coupling parameter 
space, the function $h(\lambda)$ in Eq.(\ref{free}) is expected to make
monotonic cross-over between $h(\lambda \rightarrow 0) \sim 1$ and
$h(\lambda \rightarrow \infty) \sim 3/4$. To examine whether this is the case,
in this paper, we calculate the free-energy density of the super Yang-Mills
theory in fixed-order perturbative expansion and examine its behavior as 
$\lambda$ is increased. According to our calculation, the leading-order 
${\cal O}(\lambda^2)$ correction indeed makes $h(\lambda)$ decreasing, but 
the next-order, non-analytic ${\cal O}(\lambda^3)$ correction contributes 
with (relative) opposite sign. The result up to this order then seems not to 
exhibit anticipated monotonically interpolating behavior, but it may have to 
do with truncation of the perturbation expansion to lower order terms. In an 
attempt to improve the fixed-order perturbation theory and (partially) resum 
higher-order terms, we have employed Pad\'e approximant method. Our result
shows that, once Pad\'e-improved, the function $h(\lambda)$ indeed becomes
monotonically decreasing as it is extrapolated to larger values of the 
`t Hooft coupling parameter.
Interestingly enough, the (relative) opposite sign of the 
${\cal O}(\lambda^3)$ correction  turns out precisely the origin of the 
monotonically decreasing behavior after the Pad\'e approximant is made.        

The paper is organized as follows. In section 2, we present perturbative
calculation of the free-energy density of $d=4$, ${\cal N}=4$ super Yang-Mills 
theory up to ${\cal O}(\lambda^3)$. In section 3, we improve the fixed-order
perturbative result of section 2 by Pad\'e approximant method and find smooth
interpolation plausible. In section 4, we conclude with brief discussion on
interpolation of free-energy density between $d=2$, ${\cal N}=4$ supersymmetric
$SU(N_1) \times SU(N_5)$ Yang-Mills theory and Type IIB supersring theory on 
$AdS_3$. 
Preliminary results of the present paper has been announced by one of the
authors (SJR) in \cite{nishinomiya}. Since then, while our paper is being
prepared, two independent works dealing with aspects of the calculation in 
section 2 up to ${\cal O}(\lambda^2)$ \cite{FT} and ${\cal O}(\lambda^3)$ 
\cite{vazquez} have been reported. Wherever overlapping, our results are in 
complete agreement with theirs. 

\section{Perturbative Calculation of Free-Energy}
In this section, we will be computing free-energy of four-dimensional
super Yang-Mills theory with ${\cal N}= 4$ supersymmetry and gauge group 
$G = SU(N)$ up to ${\cal O}(g_{\rm YM}^3)$ in large-$N$ limit. Throughout the
computation, we will adopt Feynman gauge and use dimensional regularization
for ultraviolet divergences. 
In component fields, the ${\cal N}=4$ super Yang-Mills theory
consists of a gauge boson, four Majorana fermions and three complex scalars 
in the adjoint representation of the gauge group $G$. 

\begin{figure}[htb]
\label{fig1}
\vspace{-1in}
\centerline{
\epsffile{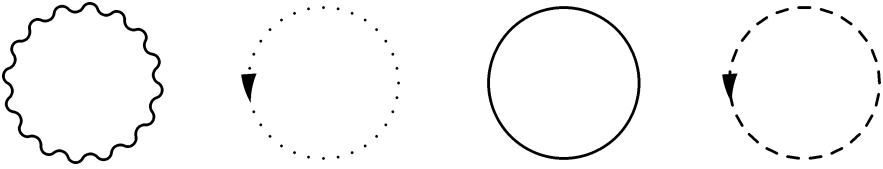}
           }
\vspace{-8.75in}
\caption{\sl One-loop Feynman diagrams contributing to free-energy $F_1$. 
Gauge bosons are represented by wiggly lines, ghosts by dotted lines, 
fermions by solid lines, and scalars by dashed lines.}
\end{figure}
\vspace{0.5cm}

At leading order, ${\cal O}(g^0_{\rm YM})$, 
the free energy density $F_1$ is simply
that of the non-interacting massless degrees of freedom and is obtained
from one-loop Feynman diagrams, as shown in 
Fig.1 for gauge bosons, ghosts, gaugino and three adjoint fermions,  and three adjoint scalars, respectively. Evaluating each diagram explicitly, 
one obtains
\bea
- F_1(G, T) = - 4 d_G \cdot\frac12 {\cal B}_0 + 2d_G \cdot\frac12 
{\cal B}_0 + 4 d_G \cdot 2 \cdot \frac12 
           {\cal F}_0 - 3 d_G \cdot 2 \cdot \frac12 {\cal B}_0 \,.
\nonumber
\eea
Here, $d_G$ denotes the dimension of the gauge group $G$, and ${\cal B}_0(T)$ 
and ${\cal F}_0(T)$ are defined as\footnote{Here, $[p]$ ($\{p\}$) represents 
that the four-momentum $p = (p_0, {\bf p})$ is bosonic (fermionic), namely,
\bea
\sum \hskip-0.5cm \int_{[p]} & \equiv & T \!\! \sum_{p_0} \, 
\int \frac{d^3 {\bf p}}{(2\pi)^3} \qquad {\rm where} \qquad p_0 = 2\pi nT \, , 
\nonumber\\
\sum \hskip-0.5cm \int_{\{p\}} & \equiv & T \!\! \sum_{p_0} 
\int\frac{d^3 {\bf p}}{(2\pi)^3} \qquad {\rm where} \qquad 
p_0 = 2 \pi (n + \frac12 T) \,.
\nonumber
\eea}
\bea
{\cal B}_0 (T) & = &\sum \hskip-0.4cm \int_{[p]} \, \ln p^2 = -\frac{\pi^2}{45}T^4\,, \nonumber \\
{\cal F}_0 (T) & = &\sum \hskip-0.4cm \int_{\{p\}} \ln p^2 = \frac78 \cdot \frac{\pi^2}{45}T^4\,.
\nonumber
\eea
Hence, summing up all the one-loop diagrams, one obtains $F_1(G, T)$ as
\bea \label{f1}
F_1 (G, T) &=& 4 d_G \left( {\cal B}_0 - {\cal F}_0 \right) 
\nonumber \\
&=& - \frac{\pi^2}{6} d_G T^4 \,.
\eea

\begin{figure}[htb]
\label{fig2}
\vspace{-1.5in}
\centerline{
\epsffile{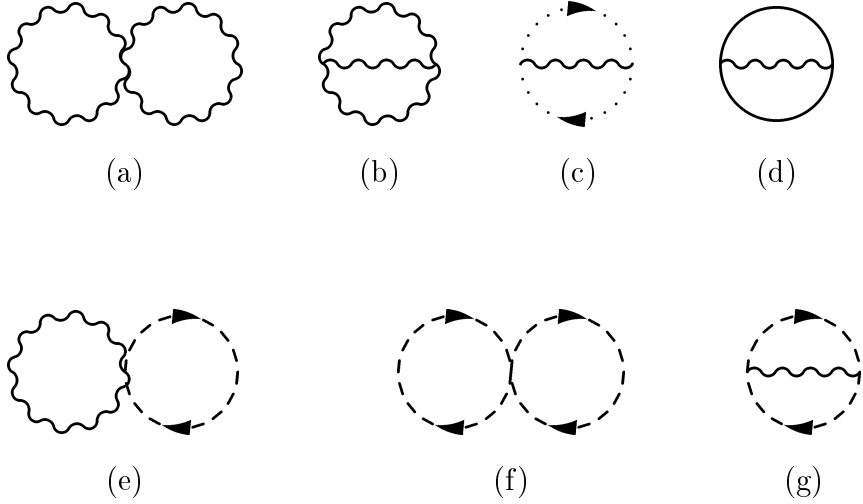}
           }
\vspace{-7.25in}
\caption{\sl Two-loop diagrams contributing to the free-energy density
$F_2$. }
\end{figure}

The contribution of the next-leading ${\cal O}(g^2_{\rm YM})$ order 
is calculated from two-loop Feynman diagrams diagrams in Fig.~\ref{fig2}. %
\vspace{0.5cm}

After a straightforward calculation, one finds that each diagram gives rise to
the following contribution:
\bea \label{f2a}
- F_{2a}&=& - 3 g^2_{\rm YM} c_A d_G {\cal B}^2 \nonumber \\
- F_{2b}&=& +\frac94 d_G g^2_{\rm YM} c_A d_G {\cal B}^2 \nonumber \\
- F_{2c}&=& -\frac14 g^2_{\rm YM} c_A d_G {\cal B}^2 \nonumber \\
- F_{2d}&=& - 16g^2_{\rm YM} c_A d_G 
         \left( {\cal B}^2 - 2{\cal B} {\cal F} \right) \nonumber \\
- F_{2e}&=& - \frac{15}{2}g^2_{\rm YM} c_A d_G {\cal B}^2 \nonumber \\
- F_{2f}&=& - 12g^2_{\rm YM} c_A d_G {\cal B}^2 \nonumber \\
- F_{2g}&=& +\frac92 g^2_{\rm YM} c_A d_G {\cal B}^2 \, .
\eea
Here, $c_A(G)$ denotes the quadratic Casimir invariant of the gauge group 
$G$ and 
\bea
{\cal B}(T) &\equiv& \sum \hskip-0.5cm \int_{[p]} \frac{1}{p^2} = + \frac{T^2}{12}\,, \nonumber \\
{\cal F}(T) &\equiv& \sum \hskip-0.5cm \int_{\{p\}} \frac{1}{p^2} = -\frac{T^2}{24}\,.
\nonumber
\eea
Adding up all the two-loop Feynman diagrams, one finds
\bea \label{f2}
F_2 (G, T) &=& 16g^2 c_A d_G \left( {\cal B} -{\cal F} \right)^2 
           \nonumber \\
           &=& \frac14 d_G \left(g^2_{\rm YM} c_A \right) T^4 \,.
\eea
Comparison with Eq.(\ref{f1}) shows that the perturbative expansion
is indeed in powers of the `t Hooft coupling parameter $
\lambda^2 \equiv  g^2_{\rm YM} c_A$.
Note that, at zero temperature, ${\cal B}_0 ={\cal F}_0$ and ${\cal B} = 
{\cal F}$ and hence bosonic and fermionic contributions in Eqs.(\ref{f1}),
(\ref{f2}) are cancelled out, manifesting the supersymmetry nonrenormalization 
theorem for the vacuum energy. For $G=SU(N)$, where $d_G=N^2-1$ and 
$c_A=N$, the results Eqs.(\ref{f1}), (\ref{f2}) are in agreement with 
\cite{FT, vazquez}.

In the fixed-order perturbative expansion, the next order would be of 
${\cal O}(g^4_{\rm YM})$, coming from three-loop Feynman diagrams. 
However, as is well-known, (a subset of) these diagrams are afflicted with 
severe infrared divergences associated with the exchange of electrostatic 
gauge bosons and scalars. 
Physically, these infrared divergences are screened by Debye plasma screening. 
The screening of gauge bosons and scalars can be taken into account by a
resummation of so-called ring diagrams to all orders \cite{grosspisarskiyaffe,
kapusta}. The
infinite-order resummation is the origin of the non-analytic contribution 
starting at order $(g^2_{\rm YM})^{3/2}$. 
An efficient way to calculate the non-analytic term is to reorganize the 
fixed-order, perturbative expansion by rewriting the Lagrangian as
\cite{AZ}
\bea
{\cal L}_{SYM} = \left( {\cal L}_{SYM} +  \frac12 M_{\rm el}^2 A_0^2 
                + m^2_{\rm el} \sum_{i=1}^3\phi_i^\dagger\phi_i \right) 
                - \left( \frac12 M_{\rm el}^2 A_0^2 
                + \frac12 m^2_{\rm el} 
                \sum_{i=1}^3 \phi_i^\dagger\phi_i\right)\,,
\nonumber
\eea
and to treat the second term as perturbative interactions. Here, 
$M_{\rm el}$ is the electric thermal mass 
for the gauge boson given by the one-loop 
self-energy $\Pi_{\mu\nu}$ at zero momentum \footnote{A nonzero 
thermal mass is generated for $A_0$, but not for $A_i$. This stems from the
fact that, after dimensional reduction over Euclidean time direction, one 
still retains $d=3$ gauge invariance.}, $ M_{\rm el}^2 \delta^{ab} = 
\Pi_{00}^{ab}(0)$, and $m_{\rm el}$ is 
the thermal mass for scalars calculated from 
$m^2_{\rm el} \delta^{ab}= \Pi^{ab} (0)$.
Self-energy diagrams in Fig.~3 can be evaluated straightforwardly in the
zero-momentum limit, much as was done in computation of Eq.~(\ref{f2a}). 
The result is 
\bea
M_{\rm el}^2&=& 2 \left( g^2_{\rm YM} c_A  \right) T^2 \,,\nonumber\\
m^2_{\rm el} &=& \left( g^2_{\rm YM} c_A \right) T^2\,.
\label{thermalmass}
\eea

\begin{figure}[htb]
\label{fig3}
\vspace{-0.9in}
\centerline{
\epsffile{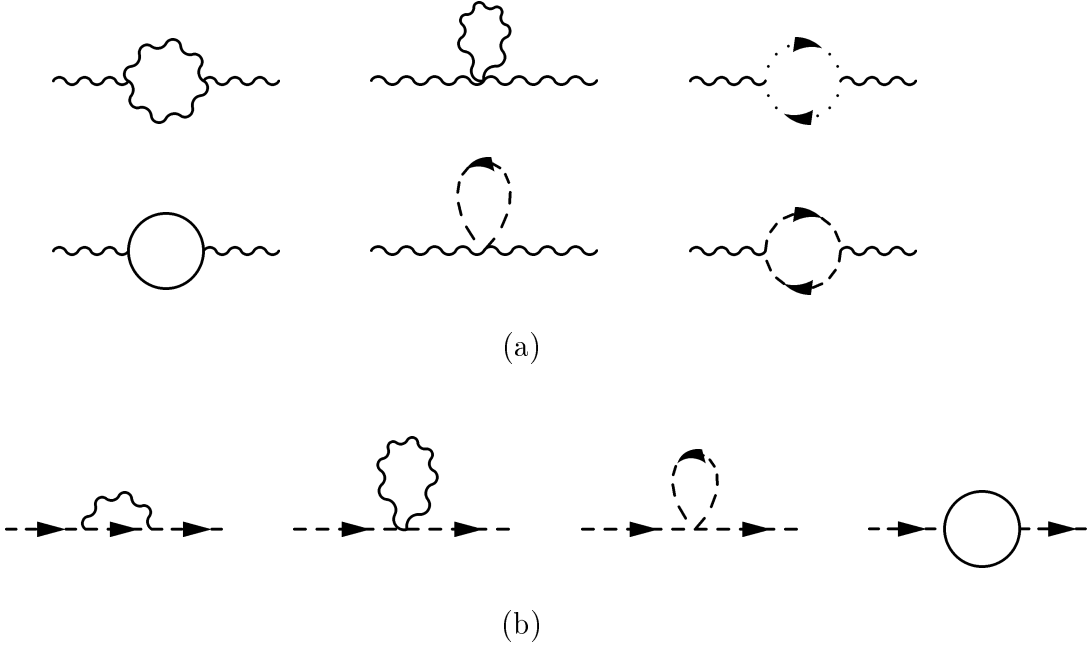}
           }
\vspace{-6.5in}
\caption{\sl One-loop self-energy diagrams for gauge bosons and scalars.
}
\end{figure}
\vspace{0.5cm}

Rather than presenting details of straightforward but tedious calculation, 
, we briefly discuss how ${\cal O}\left(g^3_{\rm YM} \right)$ contribution 
can be obtained. The free-energy density of an ideal Bose gas with mass $m$ 
is given by
\bea \label{bm}
F_{\rm free\ boson} &=& \frac12\sum \hskip-0.4cm \int_{[p]} \ln(p^2 + m^2) \,, 
                              \nonumber \\
                  &=& - \frac{\pi^2}{90}T^4 + \frac{1}{24} m^2 T^2 
                      - \frac{1}{12\pi} m^3T + {\cal O}(m^4)\,.
\eea
The first term is the well-known, black-body radiation effect of the
massless scalar particles. Its exerts a positive radiation pressure and hence
a negative contribution to the free-energy. 
The second term is the effect of nonzero mass; it contributes positively to
the free energy, as expected on physical ground. The third term, which gives 
rise to a negative correction, may be understood as representing the bosonic
nature of the particle \footnote{For a Fermi gas, there is no infrared 
divergence and hence no non-analytic contribution to the free-energy either.}. 
Its contribution is of ${\cal O}(m^3)$, thus cannot be obtained in a naive 
perturbative expansion in powers of $m^2$. If one interprets the mass $m$ as 
the thermal mass due to the Debye plasma screening,
Eq.(\ref{thermalmass}), then Eq.(\ref{bm}) may be understood as an expansion 
with respect to the coupling $(g^2_{\rm YM} c_A)^{1/2}$.
In particular, one finds that the sign of the $g^3_{\rm YM}$ term is negative.
Actually, the ${\cal O}(g^3_{\rm YM})$ term of the free energy comes entirely 
from Eq.(\ref{bm}) and there is no further correction of order $g^3_{\rm YM}$. 
The reason is simply because the thermal mass $m_{\rm el}$ is defined in such 
a way that it precisely reproduces the resummed $g^3_{\rm YM}$ effect in the
form of 
Eq.(\ref{bm}). Even though there are other $g^3_{\rm YM}$ order contributions
from individual two-loop diagrams Fig.~\ref{fig2}, they are all cancelled out
by the corresponding mass counterterm one-loop diagrams shown in 
Fig.~4. %
\begin{figure}[htb]
\label{fig4}
\vspace{-0.95in}
\centerline{
\epsffile{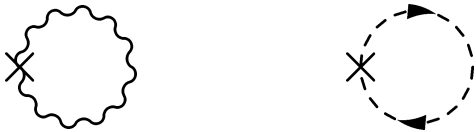}
           }
\vspace{-8.5in}
\caption{\sl Mass-counterterm diagrams due to thermal Debye masses.
}
\end{figure}%
\vspace{0.5cm}

Therefore, on general ground, the ${\cal O}(g^3_{\rm YM})$ contribution due to 
Debye plasma screening contributes negatively to the free-energy (viz. 
positive contribution to the radiation pressure), independent of the details
of the theory under consideration and can be read from Eq.(\ref{bm}) once the
thermal masses are calculated. Incidentally, extra terms of order 
$g^2_{\rm YM}$ in Eq.(\ref{bm}) are also cancelled out by the mass 
counterterms given in Fig.~4, and have no net contribution, as should be
so for consistency of the perturbative expansion. 

From the above discussion, it is now quite a simple matter to evaluate the
free-energy density $F_3$ of ${\cal O}(g^3_{\rm YM})$:
\bea \label{f3}
F_3 &=& d_G \left(-\frac{1}{12\pi} M_{\rm el}^3 T\right) 
       - 6d_G\left(-\frac{1}{12\pi} m^3_{\rm el} T\right) \,, \nonumber \\
    &=& -\frac{3+\sqrt2}{12\pi} d_G \left(g^2 c_A \right)^{3/2} T^4\, ,
\eea
where the relevant degeneracy factors are also taken into account.\footnote{
One can verify by an explicit calculation that the other terms of order 
$g^2_{\rm YM}$ and $g^3_{\rm YM}$ are all cancelled out.} 
Adding Eqs.(\ref{f1}), (\ref{f2}) and (\ref{f3})
altogether, one finally arrives at the free energy density of four-dimensional
super Yang-Mills theory with ${\cal N} = 4$ and gauge group $G$ up to 
${\cal O}(g^3_{\rm YM})$,
\bea \label{ftotal}
F= -\frac{\pi^2}{6} d_G T^4 \left\{
    1 - \frac{3}{2 \pi^2} \left(g^2_{\rm YM} c_A \right) + 
\frac{3+\sqrt2}{\pi^3} \left(g^2_{\rm YM} c_A \right)^{3/2}
    + {\cal O}\left((g^2_{\rm YM} c_A)^2\right) \right\}\,.
\eea

\section{Pad\'e Approximant of Free Energy}
The next order correction to Eq.(\ref{ftotal}) will be of order $g^4_{\rm YM}$ 
and can be calculated by considering three-loop diagrams. Moreover, a full 
account of the Debye screening at three loop order would produce terms of 
order $g^2_{\rm YM} \ln g^2_{\rm YM}$ and $g^5_{\rm YM}$. These contributions 
can be evaluated by utilizing methods developed in 
Refs.~\cite{AZ}, \cite{KZBN}. On the other hand, extending perturbative 
evaluation of the free-energy density to even higher orders breaks down 
beginning at order $g^6_{\rm YM}$. This is mainly due to another source of 
infrared problems associated with exchange of {\sl magnetostatic} gluons. 
Screening of the magnetostatic gluons cannot be achieved by resummation of 
the perturbative diagrams and, as the effect is inherently nonperturbative,
can only be treated using a reliable method such as lattice simulation 
\cite{kapusta}. As the new infrared divergence appears beginning at order
$g^6_{\rm YM}$, by scaling analysis, the {\sl magnetic} screening mass 
scale is of order $g_{\rm YM}^2 T$. 

Rather than dwelling on estimates of these effects further, one might like 
to explore implication of the result Eq.(\ref{ftotal}) to the behavior of the 
perturbative free-energy as the `t Hooft coupling parameter $\lambda$ becomes 
large. 
In particular, one would like to study if the free-energy at weak coupling
regime Eq.(\ref{ftotal}) may possibly be interpolated monotonically 
to that at strong coupling regime Eq.(\ref{free}) obtained via $AdS_5$ 
supergravity. On a first look, as $\lambda$ is increased, 
the positive sign of $g^3_{\rm YM }$ term seems to indicate that the 
result Eq.(\ref{ftotal}) at weak coupling regime behaves against any smooth 
interpolation to strong coupling regime. However, it might well have to do 
with the nature of the fixed-order perturbation theory, which truncates 
physical quantities to finite-order polynomials and induce fluctuating 
behavior. Certainly, if possible at all, one needs to improve the 
perturbative result Eq.(\ref{ftotal}) before any definite conclusion is drawn. 

Replacement of the fixed-order perturbation series by Pad\'e approximant 
\cite{pade} may
be considered as a simple tool of gaining an idea on how the series would
behave as the coupling parameter is increased. Indeed, the Pad\'e 
approximant of QCD and other field theories, 
which has been applied already to high-energy processes \cite{padeexample}
quite successfully \footnote{in the sense that Pad\'e approximant "postdicts" 
the known higher-order perturbative terms and also reduces renormalization 
scale and scheme dependences.}, seems to improve convergence of the 
perturbative free-energy in the right direction \cite{KH}. We will thus apply 
the Pad\'e's approximant to the perturbative free-energy density of 
the ${\cal N}=4$ super Yang-Mills theory and examine whether the approach 
lead to monotonic interpolation between weak and strong coupling regimes.

Let us write Eq.~(\ref{ftotal}) as
\bea \label{ratio}
R (\lambda) \equiv \frac{F( \lambda,T)}{F(\lambda=0,T)} = 1 + f_2 \lambda^2 + 
f_3 \lambda^3 + \cdots\, ,
\eea
where, from Eq.(\ref{ftotal}), $f_2 = - 3 / 2 \pi^2$ and 
$f_3 = (3 + \sqrt{2})/\pi^3$, respectively.
Note that, as it stands, $R(\lambda)$ exhibits a minimum at $\lambda = 
(-2 f_2 / 3 f_3)$, viz. $g^2_{\rm YM} N \approx 0.5$. 
The general form of the Pad\'e approximants to Eq.(\ref{ratio}) is given by
\bea
R_{[N,L]} (\lambda) 
= {1 + \sum_{n=1}^N c_n \lambda^n \over 
1 + \sum_{l=1}^L d_l \lambda^l}.
\eea
The coefficients 
$c_n$ and $d_l$ are functions of fixed-order coefficients $f_1, f_2, \cdots$.
As the two coefficients $f_1$ and $f_2$ are determined, it is possible to 
construct two distinct Pad\'e approximants, which reproduce Eq.(\ref{ratio}) 
up to order of $\lambda^3$:
\bea \label{pade}
R_{[1,2]} = \frac{1 + \xi \lambda}{1 + \xi \lambda  - f_2 \lambda^2} 
\,,\nonumber \\
R_{[2,1]} = \frac{1 + \xi \lambda + f_2 \lambda^2}{1 + \xi \lambda} \,,
\eea
respectively. Note that, as Eq.(\ref{ftotal}) contains no term of order
$\lambda$, the coefficients $c_1$ and $d_1$ are fixed to the same value:  
\bea
\xi = -\frac{f_3}{f_2} = \frac{2(3+\sqrt2)}{3\pi}\,.
\eea

\begin{figure}[htb]
\label{fig5}
\vspace{0.5in}
\centerline{
\epsfig{file=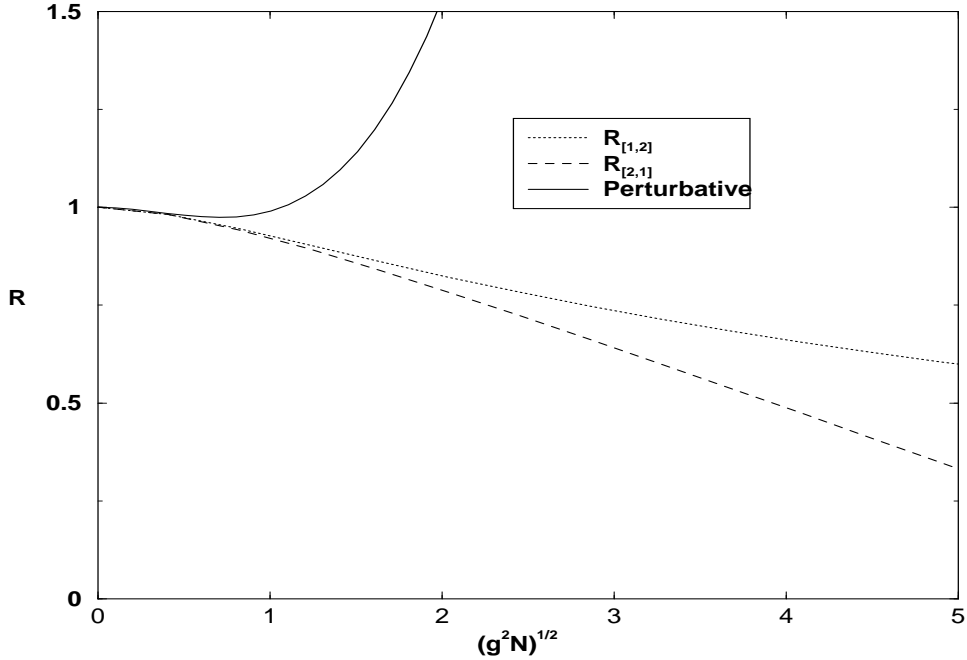, height=5in, width=3.5in, angle=-90}
           }
\vspace{0.0in}
\caption{\sl Plots of fixed-order perturbative result, and 
Pad\'e approximants $R_{[1,2]}$ and $R_{[2,1]}$.}
\end{figure}
\vspace{0.5cm}
We have plotted the two Pad\'e approximants in Fig.~5 and have made 
comparison with the perturbative result. Note that, thanks to the fact that 
$f_3$ is positive, both forms of the Pad\'e approximant show {\sl 
monotonically decreasing} behavior as the `t Hooft coupling parameter becomes 
strong. This is to be contrasted with the behavior of the fixed-order 
perturbative result, in which the strong coupling behavior is dominated by 
the $f_3 g^3_{\rm YM}$ term. Hence, within the simple Pad\'e-improvement of 
the perturbation series, we have obtained a behavior consistent with the 
strong coupling expansion drawn from the $AdS_5$ supergravity analysis and 
hence with smooth interpolation between weak and strong coupling regimes. 

Although the Pad\'e approximant would not be able to draw 
any definite conclusion regarding the strong coupling behavior,
at least one can say that the positive contribution of the $f_3 g^3_{\rm YM}$ 
term to the free-energy density is not as disastrous as it may appear for a 
smooth interpolation over the `t Hooft coupling parameter space. Explicit
calculation of higher-order $g^4_{\rm YM}$, $g^4_{\rm YM} \ln g^2_{\rm YM}$ 
and $g^5_{\rm YM}$ might be helpful in drawing more definitive conclusion on
this issue.

\section{Remark on D1+D5 and $AdS_3$ Black Hole Free-Energy}
Finally, we would like to bring up an issue regarding another interesting 
configuration exhibiting AdS/CFT correspondence: 
a bound-state of $N_1$ D1-branes and $N_5$ 
D5-branes. Effective description of the configuration
is controlled by `t Hooft coupling parameters, 
$g_{\rm YM}^2 N_1 \, , g_{\rm YM}^2 N_5$.  

At strong coupling limit, the bound-state of extremal D1-D5 brane is 
described by Type IIB string on $AdS_3 \times S_3 \times T^4$. An important
observation of \cite{gubserklebanovtseytlin} was that the geometry is not
modified even if the D1-D5 branes become {\sl non-extremal}. This is because 
of the fact that the strong coupling expansion (viz. in inverse powers 
of the two `t Hooft coupling parameters) is expressed in power-series
of Weyl curvatures and that the Weyl curvature vanishes for a product of 
$AdS_3$ and $S^3$ for equal radii of curvature. This implies that the 
free-energy density of the Schwarzschild $AdS_3$ black hole (BTZ black hole) 
is given, to all orders in strong coupling expansion, by
\bea
F= - \pi N_1 N_5 T^2 \, ,
\label{stronglimit}
\eea
viz. {\sl independent} of the `t Hooft coupling parameters.

At weak coupling limit, the D1-D5 brane configuration is described by
$d=2$ ${\cal N}=4$ supersymmetric Yang-Mills theory with gauge group $SU(N_1) 
\times SU(N_5)$ and bi-fundamental matters \cite{callanmaldacena}. Higgs-branch of the theory is 
described by four Goldstone bosons living on an effectively $N_1 N_5$ times
bigger spatial volume. Thus, thermodynamic free-energy may be calculated in
terms of sigma model of the four Goldstone bosons. At leading order
in derivative expansion, the free-energy density ought to be given by
\bea
F =4 \cdot \left( -{\pi \over 4} T^2 \right) \cdot N_1 N_5 + \cdots,
\label{weaklimit}
\eea
where the ellipses denote corrections from higher-order interaction terms of 
the sigma model. 

The leading-order term in Eq.(\ref{weaklimit}) matches
with the exact result of Eq.(\ref{stronglimit}), so if the free-energy density 
were to interpolate monotonically between weak and strong coupling regimes, 
the free-energy density Eq.(\ref{weaklimit}) should be also an exact result
of perturbative expansion in powers of `t Hooft coupling parameters. This 
implies that, in thermodynamics of $d=2$ super Yang-Mills theory, there exists 
a curious non-renormalization theorem (not related to supersymmetry). We 
suspect this may have to do with subtle noncommutativity between `t Hooft
limit, $N_1, N_5 \rightarrow \infty$, and thermodynamic infinite volume limit, 
$V \rightarrow \infty$, in two dimensional gauge theories. 
Whether this is indeed the case would serve as an interesting checkpoint to the
generality of the anticipated interpolation. Work along this direction is in 
progress and will be reported elsewhere.

\end{document}